**PAPER • OPEN ACCESS**

# Plasmonic interferences of two-particle N00N states

To cite this article: Benjamin Vest et al 2018 New J. Phys. **20** 053050

View the article online for updates and enhancements.

CrossMark

OPEN ACCESS

PAPER

# Plasmonic interferences of two-particle N00N states

Benjamin Vest[1] 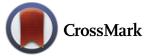, Ilan Shlesinger[1] , Marie-Christine Dheur[1], Éloïse Devaux[2], Jean-Jacques Greffet[1], Gaétan Messin[1] and François Marquier[3,4] 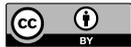

[1] Laboratoire Charles Fabry, Institut d'Optique, CNRS, Université Paris-Saclay, F-91127 Palaiseau cedex, France
[2] Institut de Science et d'Ingénierie Supramoléculaires, CNRS, Université de Strasbourg, F-67000 Strasbourg, France
[3] Laboratoire Aimé Cotton, CNRS, Université Paris-Saclay, F-91405 Orsay, France
[4] Author to whom any correspondence should be addressed.

E-mail: bvest@caltech.edu and francois.marquier@ens-paris-saclay.fr

## Abstract

Quantum plasmonics lies at the intersection between nanophotonics and quantum optics. Genuine quantum effects can be observed with non-classical states such as Fock states and with entangled states. A N00N state combines both aspects: it is a quantum superposition state of a Fock state with $N$ excitations in two spatial modes. Here we report the first observation of two-plasmon ($N = 2$) N00N state interferences using a plasmonic beamsplitter etched on a planar interface between gold and air. We analyze in detail the role of losses at the beamsplitter and during the propagation along the metal/air interface. While the intrinsic losses of the beamsplitter are responsible for the emergence of quantum nonlinear absorption, we note that N00N states decay $N$ times faster than classical states due to propagation losses.

## Introduction

Plasmon–polaritonic waves are the result of strong coupling between light and collective oscillations of electrons that propagate along a metal–dielectric interface [1]. One of their most interesting feature is the possibility to design plasmonic structures which are able to confine light in sub-diffraction limit volumes. Subwavelength confinement is at the origin of many plasmonic applications including near-field imaging, local heating, nanoantennas, and subwavelength guiding. A remarkable feature of confined fields is the fact that even with few photons, a large electric field can be generated so that the light–matter interaction is enhanced and nonlinear effects can be observed. In this limit of few photons confined to subwavelength scales, one enters the quantum nanophotonics domain. This paves the way to quantum plasmonics [2–4]. Recent experimental investigations [2] have shown that many quantum optics experiments can be reproduced using single plasmons. They demonstrate that, at some extent, solutions exist to merge the features of field confinement offered by plasmons and the quantum nature of light to perform quantum experiments. For instance, several groups reported methods to generate single plasmons, for example by using a grating to couple single photons to surface plasmons on a metal–dielectric interface [5], or by coupling a single photon emitter to a metallic nanowire [6]. The wave-particle duality of the single plasmons was checked by observing both antibunching and single-plasmon interferences in these systems. Another important step was taken with the reports of two-plasmon quantum interference between freely propagating plasmons or in plasmonic waveguides. These experiments allowed us to draw two important conclusions. First, they showed that it is possible to supply plasmonic devices with indistinguishable single plasmons, which is a fundamental requirement for any potential application in quantum information. Second, they also demonstrated that quantum features of surviving particles are well conserved even in the presence of the intrinsic losses of the metals used. Other papers focused on another fundamentally quantum feature, namely entanglement. Several papers reported that entanglement between photon states is conserved when converted into a plasmon state [5, 6], and in particular for the case of plasmonic N00N states [7]. These N00N states are of the form $|\psi\rangle_N = \frac{|N,0\rangle + e^{iN\Phi}|0,N\rangle}{\sqrt{2}}$, i.e. a superposition state where $N$ particles are in one arm and none in the other arm of a two branch interferometer. N00N states are particularly interesting when performing quantum interferences since they offer the possibility to reduce phase measurement uncertainties below the shot noise limit by a factor $\frac{1}{\sqrt{N}}$ [8, 9]. Here we report the first observation of interferences of plasmonic N00N states freely propagating along a gold–air interface and interfering on a lossy beamsplitter. In this work at the interface between plasmonics and quantum optics, we will study the interplay between quantum interferences, plasmonic confinement and losses. As opposed to quantum optics experiments in vacuum, losses are expected to play a key role in the plasmonic interferences for two reasons. First, propagation losses will be revisited for plasmonic N00N states. Second, we use a lossy beamsplitter which enables us to modify the phase difference between reflection and transmission coefficients [10–16]. It has been shown that this effect may induce nonlinear absorption. The paper is organized as follows. We first describe the experimental setup and then present the interference fringes observed when measuring detection correlations. The role of losses on the interference is discussed in the third part.

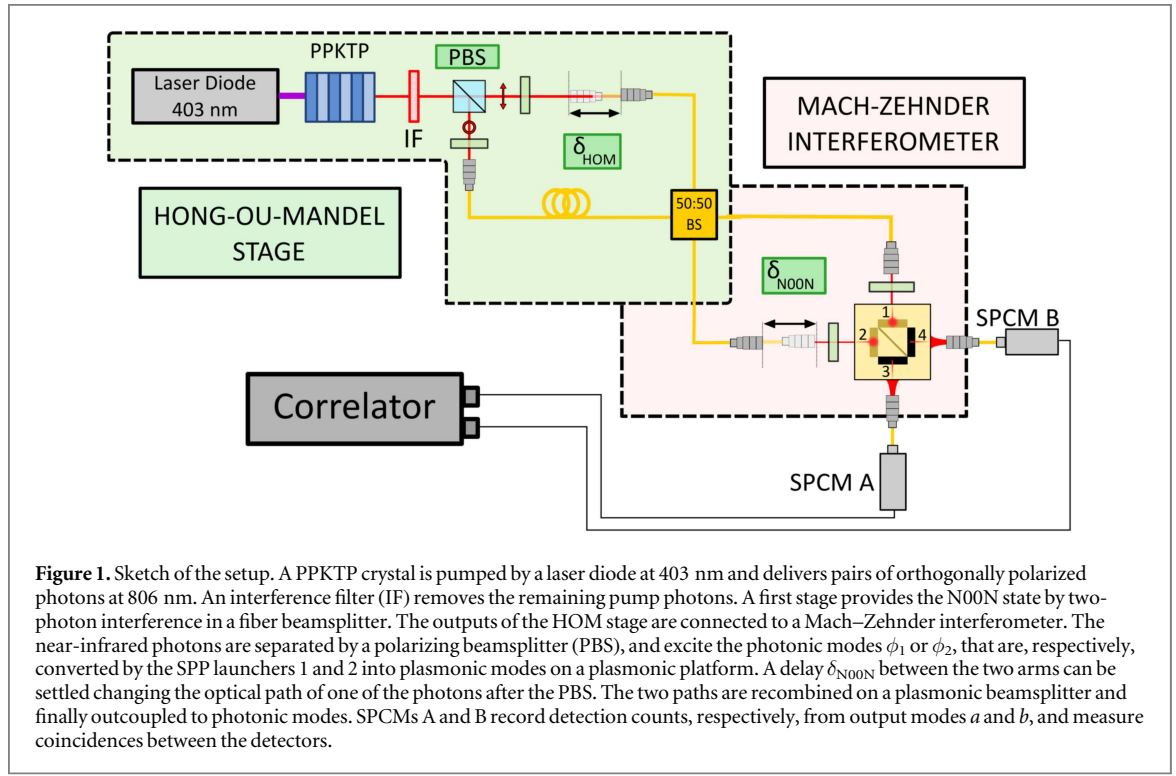

**Figure 1.** Sketch of the setup. A PPKTP crystal is pumped by a laser diode at 403 nm and delivers pairs of orthogonally polarized photons at 806 nm. An interference filter (IF) removes the remaining pump photons. A first stage provides the N00N state by two-photon interference in a fiber beamsplitter. The outputs of the HOM stage are connected to a Mach–Zehnder interferometer. The near-infrared photons are separated by a polarizing beamsplitter (PBS), and excite the photonic modes $\phi_1$ or $\phi_2$, that are, respectively, converted by the SPP launchers 1 and 2 into plasmonic modes on a plasmonic platform. A delay $\delta_{N00N}$ between the two arms can be settled changing the optical path of one of the photons after the PBS. The two paths are recombined on a plasmonic beamsplitter and finally outcoupled to photonic modes. SPCMs A and B record detection counts, respectively, from output modes *a* and *b*, and measure coincidences between the detectors.

## Methods

Let us first begin with a brief description of the experimental setup, as sketched in figure 1. It consists in the cascade of two Mach–Zehnder (MZ) interference stages. The first one uses pairs of identical photons to provide a photonic N00N state. The second MZ interferometer generates interferences of plasmonic N00N states.

Upstream of the first stage, we generate pairs of orthogonally polarized photons at $\lambda = 806$ nm thanks to a single photon down-conversion process in a periodically poled potassium titanyl phosphate (PPKTP) crystal pumped by a laser diode at 403 nm. The photons of a single pair are separated and their polarizations are aligned along the same direction before being injected in the two inputs of a fibered beamsplitter (FBS). One of the input of this FBS is mounted on a translation stage that allows us to control the relative delay $\delta_{HOM}$ between both particles, such that this entire first part of the setup reproduces a standard Hong–Ou–Mandel (HOM) experiment stage [17]. When the position $\delta_{HOM}$ is chosen so that the delay between photons is set to zero, the two particles experience coalescence and the output two-particle state is now a N00N state:

$$|\psi\rangle = \frac{|2_1, 0_2\rangle + |0_1, 2_2\rangle}{\sqrt{2}}, \qquad (1)$$

where the subscripts 1 and 2 refer to the outputs of the beamsplitter. The plasmonic N00N state interferences are observed thanks to a second hybrid MZ interferometer that introduces a second delay $\delta_{N00N}$. The outputs of the previous HOM stage are connected to the arms of the interferometer, each one being associated to an input of a plasmonic platform, where the quantum interference between plasmon states takes place. The plasmonic chips were designed by solving the electrodynamics equations with an in-house code based on the aperiodic Fourier modal method [18]. The input channels of these platforms are based on unidirectional plasmon launchers. They were optimized to efficiently couple incoming single photons into a propagating SPP on the surface [19]. The SPPs then propagate towards a surface plasmon beamsplitter (SPBS) where both plasmonic wavefunctions are recombined. Single plasmons that are not absorbed or scattered can either be transmitted or reflected at this SPBS, and then follow one of the two output paths of the chip. They finally reach the output ports of the platform consisting in slits which convert them into propagating photons.

The SPBS is intrinsically lossy. The presence of losses on the beamsplitter affects the relation between the reflection and the transmission factors ($r$ and $t$) of the beamsplitters [6, 12, 20]. More specifically, this releases constraints on the relative phase between $r$ and $t$. For a configuration where the SPBS is 50% absorbing, it is possible to modify the geometrical parameters of the SPBS to impose an arbitrary phase relation. Results are shown for a SPBS with $r = \pm t$. The consequences of the phase relation will be commented in the last section of the paper.

The relative delay $\delta_{N00N}$ can be adjusted thanks to a translation stage that modifies the photonic path followed by the N00N state before being converted into a plasmonic N00N state. After the plasmonic platform, the photons are collected by microscope objectives and injected in multimode fibers that are connected to single photon counting modules (SPCM). We record the count rates on each detector as well as the coincidence rate between both detectors within a 10 ns time window.

The coincidence detection probability can be computed from the expression of the N00N state and the beamsplitter relations linking the input modes 1 and 2 to the output modes 3 and 4. First we write the annihilation operators related to the beamsplitter modes.

$$a_3 = ta_1 + ra_2 e^{i\phi}, \tag{2}$$

$$a_4 = ra_1 + ta_2 e^{i\phi}, \tag{3}$$

where

$$\phi = \frac{2\pi \delta_{N00N}}{\lambda} \tag{4}$$

is the relative phase delay introduced between the arms of the N00N interferometer.

The coincidence detection probability can be expressed as:

$$P(1_3, 1_4) = \langle \psi | \hat{N}_3 \hat{N}_4 | \psi \rangle, \tag{5}$$

where $\hat{N}_3 = a_3^\dagger a_3$ and $\hat{N}_4 = a_4^\dagger a_4$ are the number operators of channels 3 and 4. By using the previous relations, we get:

$$P(1_3, 1_4) = 2|t|^2|r|^2(1 + \cos(2\phi)). \tag{6}$$

The coincidence detection probability thus oscillates twice as fast as the dephasing $\phi$ introduced by the path difference between the two arms. This is due to the fact that the propagation phase accumulation for a Fock state with $N$ particles, $|N\rangle$, is $N$ times higher than for a single particle state as shown in the appendix. Hence we expect to observe interference fringes when recording coincidence count rates for various delays $\delta_{N00N}$ that oscillate with half the wavelength of the incident light $\frac{\lambda}{2}$. This is the so-called phase super-resolution that is sought when dealing with N00N state interferometry. It is also relevant to point out that, as opposed to the single particle interference case [12], the N00N state interferences coincidences do not depend on the phase relation between the beamsplitter's reflection and transmission factors $r$ and $t$. Nevertheless it will have an influence on the number of particles that are absorbed as it will be discussed in the last section.

## Results

Figure 2 is a plot of the raw coincidences count rate when illuminating the interferometer with N00N states ($N = 2$) for increasing delay $\delta_{N00N}$. The signal has been integrated over 10 s. The data show fringes with a period close to 400 nm which seem to be modulated by a signal of larger wavelength. The fast Fourier transform (FFT) spectrum displayed on figure 3 indeed shows that the coincidence signal is the sum of two contributions: one main contribution at a frequency $\frac{2}{\lambda}$ and a smaller one at $\frac{1}{\lambda}$, that slightly distorts the main oscillations observed, with $\lambda = 806$ nm the wavelength of the incoming photons. A sum of two sinusoids at $\lambda = 806$ nm and $\lambda/2$ with the corresponding amplitudes given by the FFT of the data and with the phase as a free parameter have been plotted in figure 2. The two oscillations can then be understood as the expected N00N interference resulting in coincidences oscillating at twice the frequency of the incoming light which are modulated by a residual signal at the frequency of the photons. This residual perturbation of the coincidence fringes is related to a single-particle interference onto the SPBS. It can indeed be related to the initial depth of the unperfect HOM dip, that characterizes the presence in the setup of unbunched pairs of particles. We show on figure 4 the coincidence signal once the contribution at $\lambda$ has been numerically removed by applying a band-stop filter to the FFT around the SP frequency (between 0.65 $\lambda$ and 1.3 $\lambda$). This provides a clean picture of the signal, that we can fit in order to precisely estimate the frequency of the oscillation, with respect to the initial calibration that has been carried out on the single-plasmon interferences signal. We finally get a period of $397 \pm 10$ nm, which is in good agreement with the expected value around 403 nm. The observed signal exhibits a contrast around 20%, obviously much less than the ideal case of a unity visibility of the N00N fringes that can be theoretically reached. We attribute this degradation to many sources in our experimental setup: a non-perfect balancing of the SPBS and more generally of both interferometers involved in the experiment, non-optimal HOM generation of the biphoton state and the limited overlap of the modes propagating freely on the gold surface.

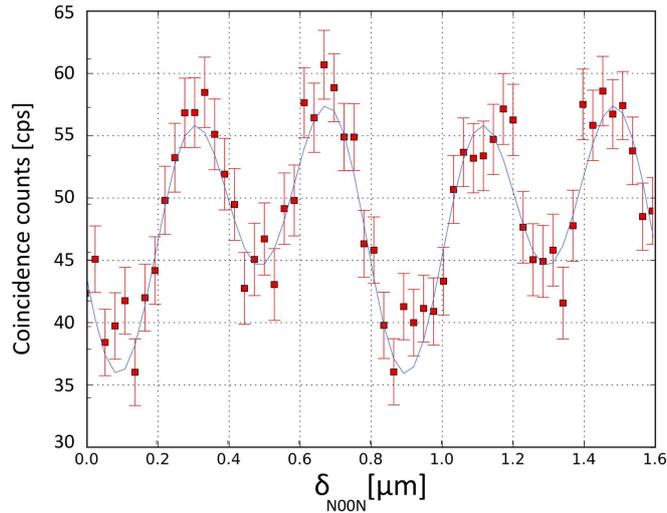

**Figure 2.** Raw data of the coincidence counts of both detectors as a function of the MZ interferometer path difference $\delta_{\text{N00N}}$. The solid line is a sum of two sinusoids at $\lambda = 806$ nm and $\lambda/2$ with fitted phases.

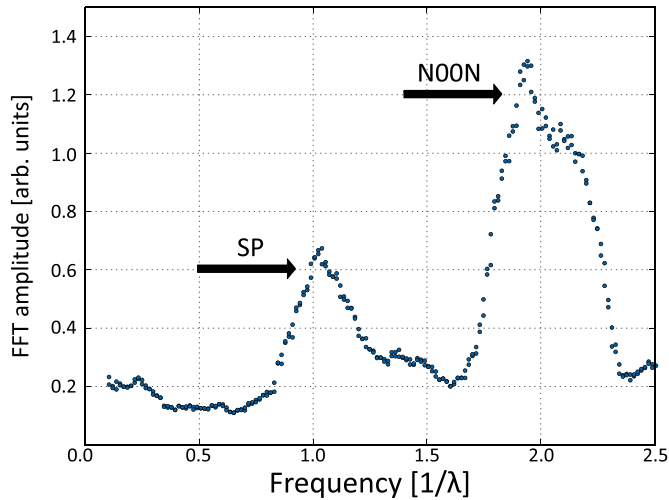

**Figure 3.** Fast Fourier transform of the raw coincidence counts of figure 2. The spectrum is plotted as a function of the inverse of the down-converted signal wavelength $\lambda = 806$ nm. The two arrows are a guide to the eye indicating the contribution of the single plasmon (SP) states with a wavenumber $1/\lambda$ and of the N00N states at twice that wavenumber.

## Discussion

Two different sources of losses are inherent to the use of plasmons during interference experiments and have different consequences when studying N00N states. First of all, there are linear losses of surface plasmon upon propagation over a distance *d*. The question is therefore what are losses for a Fock state $|N\rangle$ with $N$ particles. To get some insight into that question, we consider the phase variation $\phi = kd$ of a single particle state due to propagation over a distance *d*, with $k = \frac{2\pi}{\lambda}$ the wavevector. As previously mentioned, this becomes $\phi_{\text{N00N}} = Nkd$ for the state $|N\rangle$. Introducing losses by considering that the wavevector $k = \frac{2\pi}{\lambda}$ is complex $k = k' + ik''$, we expect to observe a decay length of the state $\delta = \frac{1}{2Nk''}$ $N$ times smaller than for a single particle state. We prove this result in the appendix using the usual beamsplitter model for propagation in a lossy medium. The result can be interpreted with a naive picture: the transmission probability of each particle through $d$ is given by $\exp(-d/\delta)$ so that the transmission probability of the $N$ particles is given by $\exp(-Nd/\delta)$.

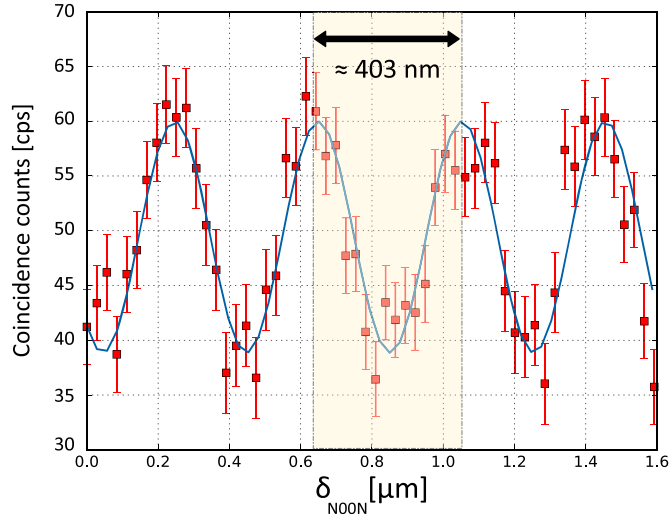

**Figure 4.** Coincidence counts (squares) with error bars as a function of the MZ interferometer path difference $\delta_{\text{N00N}}$ when the contribution of SP states has been numerically filtered. The solid line is a sinusoid at twice the frequency than the incident light (refer to figure 5).

The second source of losses is due to the plasmonic beamsplitter itself. As experimentally shown in a previous paper, the presence of losses in the beamsplitter, i.e. $|t|^2 + |r|^2 < 1$, allows us to modify the phase relation between the reflection and transmission factor. While a lossless beamsplitter imposes a constrained phase relationship between $r$ and $t$, namely $r = \pm it$, a lossy beamsplitter can be designed to exhibit any different phase relation. It has been theoretically and experimentally shown that new quantum effects arise for the case $r = \pm t$ [10, 13, 14]. Depending on the phase relation of the input N00N state, one can deterministically obtain either a single photon state or a mixture of zero and two photon states at the output. This phenomenon has been coined quantum nonlinear absorption and has been observed in a recent experiment [14]. In this setup, evidence of being in such a coherent absorption regime is given by the reminiscent single plasmon oscillations shown in figure 5. Indeed, the observed phase shift between signals from SPCMs A and B is close to 0, and as it was shown in a previous article [12], it is the direct consequence of the phase relation $r = \pm t$. The in-phase evolution of the signals can be interpreted as the successive preferential transmission or absorption of single particles. We can therefore assume that the evolution of the N00N interference signal follows an analogous scheme: when the maxima are reached, the output state mix preferentially contains two-particle states (thus increasing the number of coincidences). When the minima are reached, one gets more single particle states, thus reducing coincidence counts.

In summary, we have observed quantum interference of a plasmonic $N = 2$ N00N state freely propagating on a plasmonic platform. This experiment is a further demonstration of the preservation of advanced quantum behavior of photons when converted into plasmons and vice versa, even in experimental conditions that can generally be considered as highly detrimental for quantum systems, such as the use of lossy materials and/or structures.

## Acknowledgments

We thank A Browaeys and P Grangier for fruitful discussions. The research was supported by a DGA-MRIS (Direction Générale de l'Armement Mission Recherche et Innovation Scientifique) scholarship, by the SAFRAN-IOGS chair on Ultimate Photonics and by a public grant overseen by the French National Research Agency (ANR) as part of the Investissements d'Avenir Programme (reference: ANR-10-LABX-0035, Labex NanoSaclay). J-JG acknowledges the support of Institut Universitaire de France.

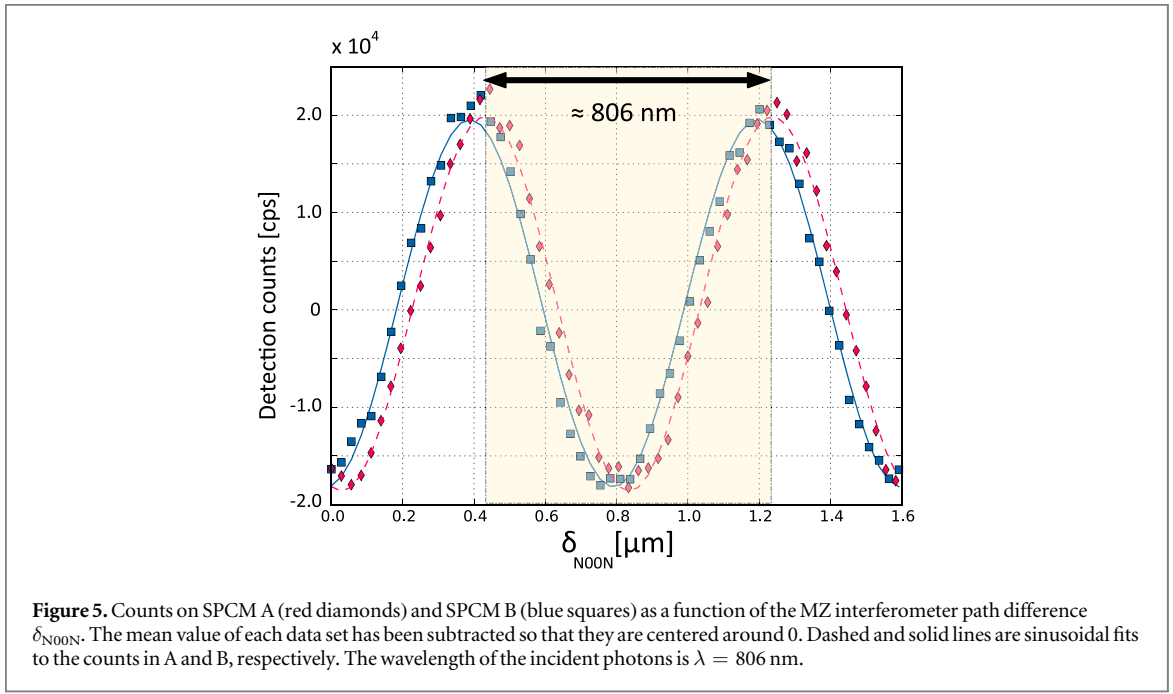

**Figure 5.** Counts on SPCM A (red diamonds) and SPCM B (blue squares) as a function of the MZ interferometer path difference $\delta_{N00N}$. The mean value of each data set has been subtracted so that they are centered around 0. Dashed and solid lines are sinusoidal fits to the counts in A and B, respectively. The wavelength of the incident photons is $\lambda = 806$ nm.

## Appendix A. Materials

### Detection method

All the photons in these experiments were sent to fibered SPCMs, which deliver transistor-transistor logic pulses. SPCMs A and B are PerkinElmer modules (SPCM AQRH-14). To count the correlations between the SPCMs A and B pulses, we used a PXI Express system from National Instruments (NI). The NI system is composed of a PXIe-1073 chassis on which NI FlexRIO materials are plugged: a field programmable gate array (FPGA) chip (NI PXIe-7961R) and an adapter module at 100 MHz (NI 6581). The FPGA technology allows us to change the setting of the acquisition by simply programming the FPGA chip to whatever set of experiments we want to conduct. A rising edge from SPCM A or B triggers the detection of another rising edge, respectively, on channel B or A at specific delays. Counting rates and coincidences between channels A and B are registered. The resolution of the detection system is mainly ruled by the acquisition board frequency clock at 100 MHz, which corresponds to a time resolution of 10 ns.

### Photon pair source

The photon pairs source is based on parametric down-conversion. A potassium titanyl phosphate crystal (PPKTP crystal from Raicol) crystal is pumped at 403 nm by a tunable laser diode (Toptica). It delivers a 38 mW powered-beam, focused in the crystal by a 300 mm focal-length plano-convex lens. The waist in the crystal is estimated to be 60 $\mu$m. The crystal generates pairs of orthogonally polarized photons at 806 nm. The waist in the crystal is conjugated to infinity with a 100 mm focal-length plano-convex lens, and the red photons emerging from the crystal are separated in polarization by a polarizing beamsplitter (PBS) cube (Fichou Optics). We remove the remaining pumping signal with a 1 nm-spanned interference filter (IF) from AHF (FF01-810/10).

### Coupling to the platform and collection of the output photons

Each photon of a pair is then coupled to one input mode of a polarization maintaining monomode FBS (Thorlabs). The output state is outcoupled via Long Working Distance M Plan Semi-Apochromat microscope objectives (LMPLFLN-20X BD, Olympus) and sent to two different inputs of a PBS (Fichou Optics) with orthogonal polarizations. They leave the cube by the same output port and were focused with a 10× microscope objective (Olympus) on the plasmonic sample. The plasmonic sample is mounted on a solid immersion lens. The surface plasmons propagating on the chip leave the sample by two orthogonal output slits. The conversion of the SPPs back to photons via the slits leads to two different directions of propagation in free space. The photons from the output ports are collected from the rear side of the sample using mirrors and a 75 mm focal-length lens for each output. The output modes are then conjugated to multimode fibers via a 10× microscope objective (Olympus), which are connected to the SPCMs.

**Plasmonic platform sample fabrication**

We deposited 300 nm thick gold films on clean glass substrates by e-beam evaporation (ME300 Plassys system) at a pressure of $2 \times 10^{-6}$ mbar and at a rate of $0.5$ nm s$^{-1}$. The rms roughness is 1 nm. The films were then loaded in a crossbeam Zeiss Auriga system and milled by a focused ion beam at low current (20 pA), except for the large slits used to decouple plasmons for propagating light that were milled at 600 pA.

## Appendix B. Calibration and residual count rate estimation

Both SPCM A and B exhibit some residual oscillation from a single-plasmon interference component. The data is fitted in order to get a calibration of the setup for the down-converted signal wavelength at $\lambda = 806$ nm with respect to the relative motion of the translation stage. Figure 5 is a plot of the count rates on SPCMs A and B when increasing the path difference $\delta_{\text{N00N}}$ between both arms of the second MZ interferometer. Those oscillations are due to single-plasmon interferences, associated to particles who did not bunch into a biphoton state before entering the MZ, either because their partner was lost, or because of the unperfect generation of the N00N state at the first stage beamsplitter. Hence some detection events can be considered as the consequence of single particles entering the interferometer through the FBS. When the delay is close to zero, the particles can propagate along indistinguishable paths through the setup and quantum paths interference occur: the count rates exhibit a small component oscillating at the frequency of the incident light $\lambda = 806$ nm. The amplitude of the single-plasmon interference oscillations can be linked to the initial depth of the unperfect N00N state generation from the HOM stage and from various unbalance sources in the setup. It can also be shown, as noted in [14], that unbunched particles leading to single-plasmon interferences can also formally contribute to the coincidence signal at $\frac{2}{\lambda}$, with the help of their twin particles. On balance, this does not affect the amplitude of the N00N oscillations by more than few percents.

## Appendix C. Propagation losses

Phase shift and losses through propagation are described by elementary unitary beamsplitters where the reflected signal accounts for the losses and the remaining signal is the one transmitted [21]. The reflection and transmission coefficients of each beamsplitter satisfy the relations:

$$|t|^2 + |r|^2 = 1, \tag{7}$$

$$tr^* + t^*r = 0. \tag{8}$$

The field operators of the output arms 3 and 4 of the beamsplitter can be written as a function of the field operator of the input arms 1 and 2 as:

$$\hat{a}_3 = t\hat{a}_1 + r\hat{a}_2, \tag{9}$$

$$\hat{a}_4 = r\hat{a}_1 + t\hat{a}_2, \tag{10}$$

which gives:

$$\hat{a}_1^\dagger = t\hat{a}_3^\dagger + r\hat{a}_4^\dagger, \tag{11}$$

$$\hat{a}_2^\dagger = r\hat{a}_3^\dagger + t\hat{a}_4^\dagger. \tag{12}$$

When considering propagation in a lossy medium between $x$ and $x + \delta_x$ one obtains [21]:

$$t = \exp(ik\delta_x) = \exp(i(k' + ik'')\delta_x), \tag{13}$$

where $k$ is the complex wavevector with the real part and the imaginary part describing, respectively, the phase shift and the attenuation. Let us now consider a Fock state $|N_1, 0_2\rangle$, where subscripts denote the arms of the beamsplitter, as the input state $|\Psi_{\text{in}}\rangle$. This state is written:

$$|\Psi_{\text{in}}\rangle = |N_1; 0_2\rangle = \hat{a}_1^{\dagger N}|0_1; 0_2\rangle. \tag{14}$$

When using equation (11) to replace $\hat{a}_1^\dagger$ one obtains the corresponding output state:

$$|\Psi_{\text{out}}\rangle = (t\hat{a}_3^\dagger + r\hat{a}_4^\dagger)^N|0_3; 0_4\rangle \tag{15}$$

$$= \sum_k^N \binom{N}{k} t^k r^{N-k} (\hat{a}_3^\dagger)^k (\hat{a}_4^\dagger)^{N-k}|0_3; 0_4\rangle. \tag{16}$$

The $N$ photon state is only preserved in the output arm 3 when the $N$ photons are transmitted which corresponds to the term:

$$|N_3; 0\rangle = t^N (\hat{a}_3^\dagger)^N |0_3; 0_4\rangle. \tag{17}$$

This equation together with equation (13) give the amplitude factor for the propagation of a *N* particle state through a distance $\delta_x$:

$$t^N = \exp(Nik\delta_x) = \exp(i(Nk' + iNk'')\delta_x) \tag{18}$$

showing that both the phase shift and the attenuation are *N* times larger.

## ORCID iDs

Benjamin Vest  https://orcid.org/0000-0003-3640-4560
Ilan Shlesinger  https://orcid.org/0000-0002-9328-1926
François Marquier  https://orcid.org/0000-0003-3118-1150